# Optimizing Metro Station Locations and Line Layouts in Selangor using Genetic Algorithm Approach: Technical Report

Hasna Lammaihri, Marwa Erramla
School of Industrial Management,
Mohammed VI Polytechnique University
Lot 660, Hay Moulay Rachid Ben Guerir
4315 Morocco

Norjihan Abdul Ghani
Department of Information Systems,
Faculty of Computer Science and Information Technology,
Universiti Malaya,
50603, Kuala Lumpur, Malaysia

## Abstract

This report presents an approach for optimizing metro station locations and line layouts in the area of Selangor, located in Malaysia. The project utilized the genetic algorithm in identifying the locations and lines layout. With Selangor's population projected to reach 7.3 million by 2024, the existing transport infrastructure is under increasing strain. This project addresses this challenge by optimizing the metro network to effectively cover the expanding population and minimize travel times. We employed a genetic algorithm to achieve these objectives, focusing on both the strategic placement of metro stations and the efficient layout of metro lines.

## 1    Introduction

Selangor is undergoing rapid urban growth, which imposes significant demands on its public transportation system. To accommodate this growth and ensure efficient urban mobility, it is essential to develop a well- designed metro network. The existing transport infrastructure is struggling to keep up with the increasing population, necessitating a comprehensive approach to metro network optimization. This project addresses the complexity of planning a metro system, focusing on optimizing the layout of metro stations and the network of lines connecting them. The main challenge addressed in this project is the design of metro station locations and line layouts that efficiently serve Selangor's growing population while minimizing travel times. The goal is to develop an optimized metro network that enhances coverage and reduces travel times for passengers. The proposed solution able to create an efficient and cost-effective metro network that meets the transportation demands of inhabitants. Therefore, the purpose of this project is to optimize the metro network in Selangor to better serve its expanding population and minimize travel times. According to Blanco, et al., (2020), the planning process includes several phases that usually are sequentially executed in the following order:

- Network design, where the stations, links and routes of the lines are established.
- Line planning, specifying the frequency and the capacity of the vehicles used in each line. (line concept)
- Timetabling, defining the arrival/departure times, and
- Scheduling, in which vehicles and/or crews are planned

Most recent studies using GAs in transit network problem can be found for optimization of route network design (Erban, 2021). Based on this, for this paper purpose, there are two key points here:

i. Metro Network Planning: The process is complex due to the large amount of required input data from various domains and the significant budget constraints.

ii. Genetic Algorithm: it employs a genetic algorithm for designing metro networks, which helps in handling the computational complexity and finding near-optimal solutions efficiently.

The study presented in this paper covers two stages of metro system design; Stage 1, the optimal locations of metro stations were found, based on population density and transportation demands and Stage 2; optimal layout of metro lines was designed as proposed by Król and Król (2019).

## 2 Case Study Description

The initial step in this study involved gathering and processing data relevant to the metro network design. Selangor, as shown in Figure 1 is part of the Klang Valley metropolitan area, faces significant transportation challenges due to its dense urban structure, extensive bus and tram lines, low-frequency railway network, and congested roads. This preliminary design outlines a metro system aimed at improving connectivity, reducing traffic congestion, and promoting sustainable urban development in Selangor.

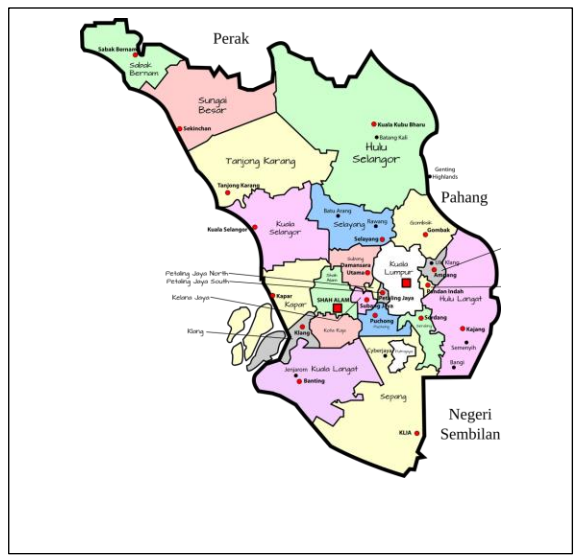

Figure 1: Map of Selangor

*Geospatial data*: Collected data includes detailed geographical boundaries of the target region, such as Selangor as shown in Figure 2. GeoJSON files were used to define the boundaries of districts and municipalities.

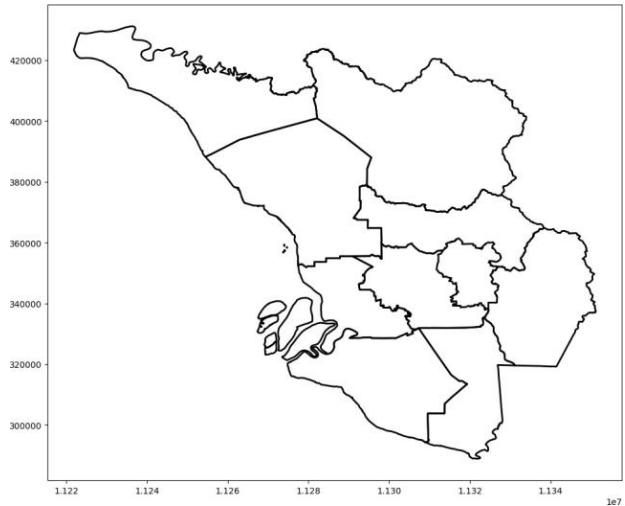

Figure 2: Geospatial Visualization of Selangor's Boundaries

***Population density:*** Population data is essential to identify areas of high transportation demand. This data was obtained from DOSM Data Catalogue (DOSM, 2024). However, we have further refined the data. Figure 3 below shows the final density map:

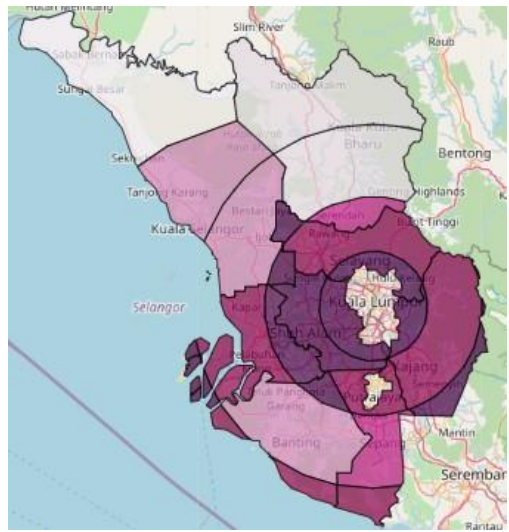

Figure 3: Population Density Map of Selangor

***Transportation demand generators***: Key points of interest like business districts, residential areas, airports, schools, and other major hubs were identified as shown in Figure 4. These demand generators significantly impact the placement of metro stations.

| Centre | Number of daily visitors | Latitude | Longitude |
|---|---|---|---|
| Batu Caves | 5000 | 3.2379 | 101.6841 |
| Sunway Lagoon | 8000 | 3.0713 | 101.6078 |
| i-City | 3000 | 3.0675 | 101.4856 |
| Zoo Negara | 2500 | 3.2133 | 101.7617 |
| Bukit Melawati | 1000 | 3.3463 | 101.2570 |
| Royal Selangor Visitor Centre | 500 | 3.1990 | 101.7055 |
| Kuala Selangor Nature Park | 400 | 3.3421 | 101.2417 |
| Shah Alam Lake Gardens | 2000 | 3.0730 | 101.5183 |
| Sekinchan | 700 | 3.5167 | 101.0994 |
| The Mines | 2500 | 3.0349 | 101.7161 |
| Sunway Pyramid | 50000 | 3.0731 | 101.6071 |
| 1 Utama | 40000 | 3.1507 | 101.6151 |
| Cyberjaya | 100000 | 2.9223 | 101.6509 |
| Petaling Jaya | 200000 | 3.1073 | 101.6067 |

Figure 4: Main Attraction Centers of Selangor

## 3 Genetic Algorithm Setup

After collecting and preprocessing the data, the next step involved setting up the genetic algorithm, which was chosen due to its ability to efficiently search large solution spaces for near-optimal designs.

***Initial population generation:*** Firstly, the algorithm necessitates predefined numbers of stations and lines, so we need to determine how many stations and lines to implement in Selangor. To find the optimal numbers, we conducted a benchmark and created plots to analyze the relationship between population and the number of metro stations, as well as between population and the number of metro lines as shown in Figure 5 and 6.

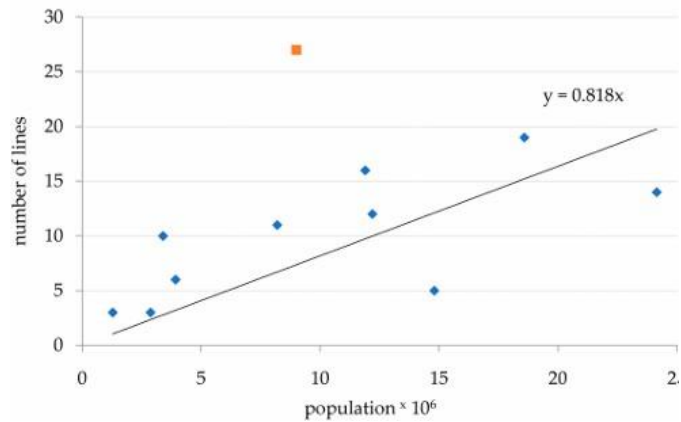

Figure 5: Dependency between the Number of Metro Lines and the Number of the Stations

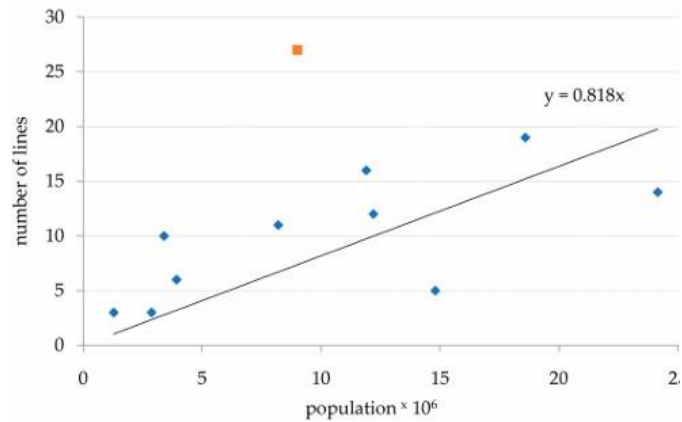

Figure 6: Dependency between the Number of Metro Lines and the Number of the Lines

## Genetic algorithm overview

The genetic algorithm (GA) is inspired by biological evolution, using processes like selection, crossover, and mutation to optimize solutions. In the context of optimizing metro station locations, GA helps identify the most effective positions by evolving candidate solutions over generations. GAs exhibit robustness and flexibility in exploring wide search spaces and gradually converging to optimal or near-optimal solutions, making them particularly suitable for this problem (Biao, et al., 2024). Figure 7 shows a diagram illustrating the genetic algorithm.

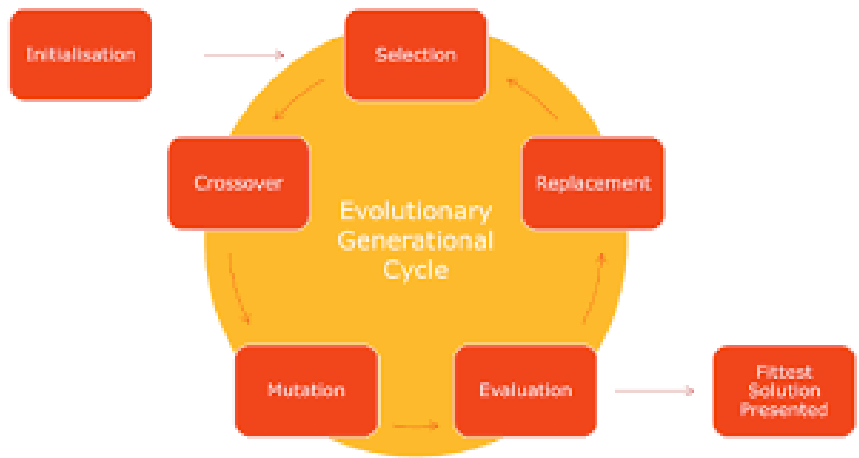

Figure 7: Genetic Algorithm Scheme

The first stage of genetic algorithm approach is initialization where we generate a population, here we for each stage we generate 2 population for stations and for lines, each population contains individuals.

An **individual** is a single possible solution in your search space. In your metro station optimization project, an individual might be a set of coordinates representing possible metro station locations. Each individual is evaluated using a **fitness function** that determines how good that solution is (e.g., how well the metro stations serve the population).

Here are the two populations that were generated:

- Station placement: The first population of potential metro networks was generated by placing stations randomly within the study area while considering population density and geographic constraints. Each individual in the population represented a different set of station placements as shown in Figure 8.

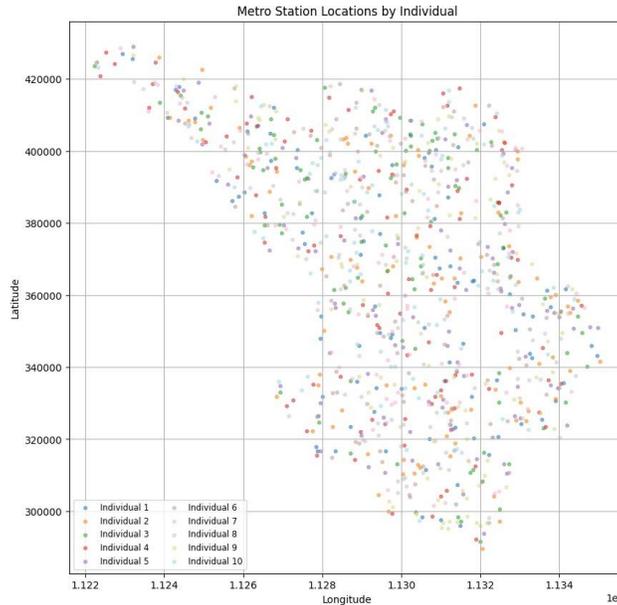

Figure 8: Population of Randomly Generated Metro Stations

- Metro line layout: After station placement, initial metro line configurations were created. Each line connected a subset of the stations, and each individual in the population had five lines. This provides a glimpse into the generated population of metro lines as shown in Figure 9

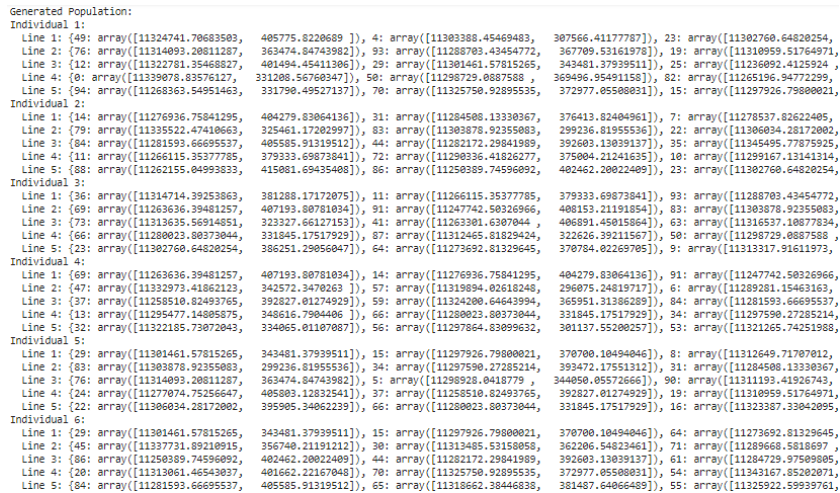

Figure 9: Population of Randomly Generated Metro Lines

This population undergoes an optimization process that is repeated across generations.

A **generation** represents one complete cycle of creating a new population from the current one. In each generation, a set of individuals (potential solutions) undergoes processes like **selection**, **crossover**, and **mutation** to produce new individuals for the next generation. After each generation, the new population becomes the current population for the next cycle.

**Optimization process**

According to Kaleybar, et al. (2023), optimization techniques can play a significant role in improving the efficiency, safety, and reliability of the RSs. By using optimization techniques, it is possible to reduce operating costs, increase capacity, improve the accuracy of train schedules, and minimize the risk of accidents

There are two stages involved:

i. Station location optimization: Genetic algorithms are used to maximize the number of people served by each station. Key factors such as proximity to demand generators and population density are taken into account.
ii. Metro line layout optimization: In the second stage, the layout of metro lines connecting the stations is optimized to minimize travel times. The line design prioritizes network coherence, ensuring that all stations are reachable.

Passing an individual to the next generation means that after the current generation's selection, crossover, and mutation processes, some individuals (solutions) are selected to form the population for the next generation.

***Selection***: The best-performing individuals (based on fitness) are chosen to have "offspring" in the next generation. These offspring are created using genetic operations (crossover and mutation) based on the selected individuals. (Refer Figure 10).

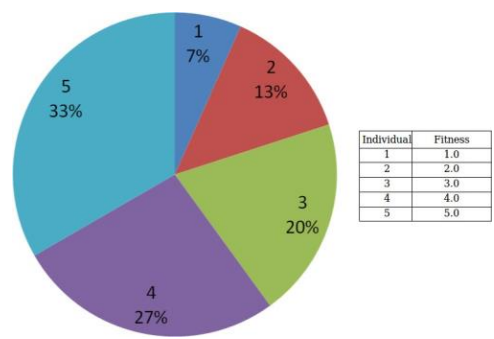

Figure 10: Roulette-wheel-selection

***Crossover (Recombination):*** Two selected individuals (parents) combine parts of their genetic information to create new individuals (children). This process mimics reproduction in nature. For example, in your metro stations problem, crossover might mean taking part of one parent's metro station locations and combining it with part of another parent's locations to create new station configurations. (Refer Figure 11).

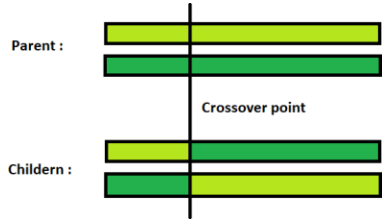

Figure 11: Crossover Scheme

*Mutation:* After crossover, a small random change is applied to some of the individuals to introduce variability. This prevents the population from becoming too similar and helps the algorithm explore new potential solutions. (Refer Figure 12).

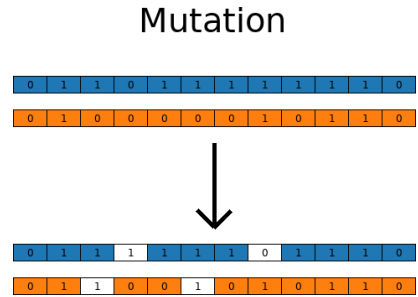

Figure 12: Mutation Example

In this report, mutation involves shifting the position of a randomly selected metro station to explore new configurations. This is achieved by adding random values to the coordinates of the selected station. Specifically, a normal distribution with a mean of zero and a standard deviation, which is a parameter of the program, generates these random values. Mathematically, if a station at coordinates ($x_i$, $y_i$) is selected for mutation, the new coordinates are calculated as:

$$(x\prime_i, y\prime_i) = (x_i + \Delta x, y_i + \Delta y)$$

where $\Delta x$ and $\Delta y$ are random values drawn from a normal distribution. This mutation technique allows for the introduction of variability in the station locations, aiding the genetic algorithm in finding more effective and diverse solutions for the metro network design. And for the metro lines, six types of mutations are used to modify the metro lines:

(a) Exchange of the position of two stations within a line.
(b) Reversion of the order of a sub-series of stations within a line.
(c) Exchange of a station between two lines.
(d) Transfer of a station from one line to another.
(e) Removal of a station from a line.
(f) Addition of a station to a line.

***Elitism*** (Optional): With elitism, the best individuals from the current generation are **directly copied** into the next generation without undergoing crossover or mutation. This ensures that the best solutions are preserved and not lost due to random changes.

***New population***: Once selection, crossover, mutation, and possibly elitism have been applied, a new population of individuals is formed. This population replaces the old one, and the process repeats for a certain number of generations. In this case, 10 generations were used, as this was the point at which the solutions nearly converged and approached optimality, while also addressing certain computational constraints.

# 4 Applying the Genetic Algorithm to Identify Optimal Locations for Metro Stations

**Fitness function**

The total coverage $f$ is given by:

$$f_{obj}^1 = \sum_i \left( \int_A e^{-\frac{r_i^2}{\sigma^2}} \rho \, dA \right) + \sum_i \sum_j \left( e^{-\frac{r_{ij}^2}{\sigma^2}} g_j \right)$$

Where:

$r_i$ is the distance from the *i*-th station to the centroid of the district,

$\rho$ is the local population density,

$\sigma$ is the standard deviation of the Gaussian function,

$A$ is the considered area,

$r_{ij}$ is the distance between the *j*-th generator and the *i*-th station,

$g_j$ is the number of people related to the *j*-th generator.

This function can be broken down into two parts:

i. Coverage from District Population Densities:

$$\sum_i \left( \int_A e^{-\frac{r_i^2}{\sigma^2}} \rho \, dA \right)$$

Here, A represents an integral over the area A of each district. $e^{-r_i^2/\sigma^2}$ is the Gaussian function representing the influence of the station on the district. $\rho$ is the population density of the district. The integral calculates the coverage over the entire area of each district.

ii. Coverage from generators:

$$\sum_i \sum_j \left( e^{-\frac{r_i^2}{\sigma^2}} g_j \right)$$

This will sum over all stations *i* and all generators *j*. $e^{\sigma^2}$ is the Gaussian function representing the influence of the station on the generator. $g_j$ is the population count at generator *j*.

## Criteria of Station Locations

The passage outlines a model used for determining the optimal locations of metro stations to meet the transportation demands of the inhabitants. Here's a breakdown of the key points:

1. **Station Location and Transportation Demand**: A metro station's location must satisfy the transportation needs of the population in its vicinity. The demand for transportation is assumed to be proportional to the number of people living in an area.

2. **Input Data**: The primary input for this model is a map showing population density across the region.

3. **Simplified Approach**: In this model, a rough approach is used that disregards the street layout and existing mobility options of inhabitants. The model assumes that passenger satisfaction depends on the Euclidean distance (straight-line distance) to the nearest metro station, with circular catchment areas for each station.

4. **Satisfaction Model for Residents**: The number of inhabitants satisfied with the location of a particular metro station *i* is expressed by an integral equation that considers the distance from the station ($r_i$), the local population density ($\rho$), a standard deviation parameter ($\sigma$) that determines the acceptable distance, and the area (*A*) being considered.

5. **Point Generators of Demand**: The model also accounts for point generators of transportation demand, such as stadiums, airports, and factories. The number of people related to these generators who are satisfied by a particular metro station *i* is calculated based on the distance from the generator to the station ($r_{ij}$) and the number of people associated with the generator ($g_j$).

6. **Parameter $\sigma$**: The value of $\sigma$ is crucial as it defines the distance passengers find acceptable for reaching a metro station. Different studies have suggested varying values, generally ranging between 400 meters and 3000 meters, depending on the mode of access (walking, Park & Ride systems, etc.).

7. **Optimization Objective**: The objective of the optimization process is to vary the locations of all stations to maximize the total number of people whose transportation needs are satisfied. This is represented by the objective function $f_{1obj}$ which sums the number of satisfied inhabitants from both the general population and the point generators.

8. **Final Output**: After optimization, the best locations for all stations are identified, and the number of potentially serviced people is calculated for each station, denoted as $s_i$, where

$$s_i = n_i + \sum_j n_{g_{ij}}$$

Based on the criteria, it describes a model that optimizes the placement of metro stations based on population density and specific demand points to maximize the number of people whose transportation needs are satisfied. Table 1 explains possible assumptions that need to be considered when designing metro station locations.

Table 1: Important Assumptions for Designing Metro Station Locations

| Assumption | Description |
| --- | --- |
| Proportional Transportation Demand | Transportation demand is directly proportional to population density. More populated areas have higher demands. |
| Euclidean Distance Criterion | Passenger satisfaction depends on the straight-line distance to the nearest metro station. Shorter distances increase satisfaction. |
| Circular Catchment Areas | Each station serves a circular area around it. People within this radius are considered served by the station. |
| Point Generators Consideration | Key locations like stadiums, airports, and factories are considered separately in the model. |

**Result**

The following Figure 13 illustrates the optimal locations for the metro stations, highlighting how these positions achieve the best configuration according to our optimization algorithm.

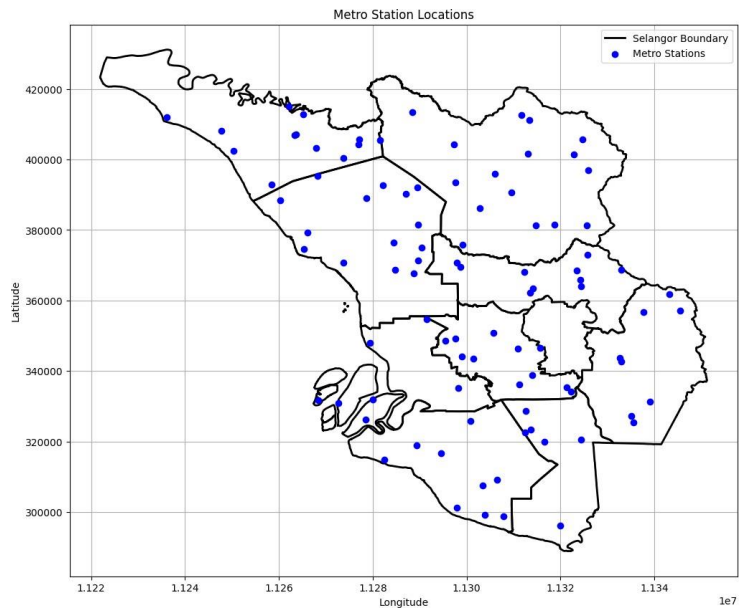

Figure 13: Locations for the metro stations.

# 5 Applying the Genetic Algorithm to Identify Optimal Layout of Lines Connecting the Stations

The goal is to design a metro network by optimizing the layout of metro lines, where:

- Each line $L_i$ is a series of stations.
- A station can belong to multiple lines, but it cannot form loops within a line.
- Any station should be reachable from any other station.
- Various mutation operations are used to modify the lines, and cross-over operations allow for genetic diversity between different possible layouts.
- here is the fitness function in this case :

$$f_{obj}^2 = \sum_{i,j} t_{ij} = \sum_{i,j} d_{ij}\left(s_i + s_j\right) \to min$$

**Result**

The following Figure 14 demonstrates the optimal configurations of the metro lines, illustrating how these configurations provide the most efficient layout according to our optimization algorithm.

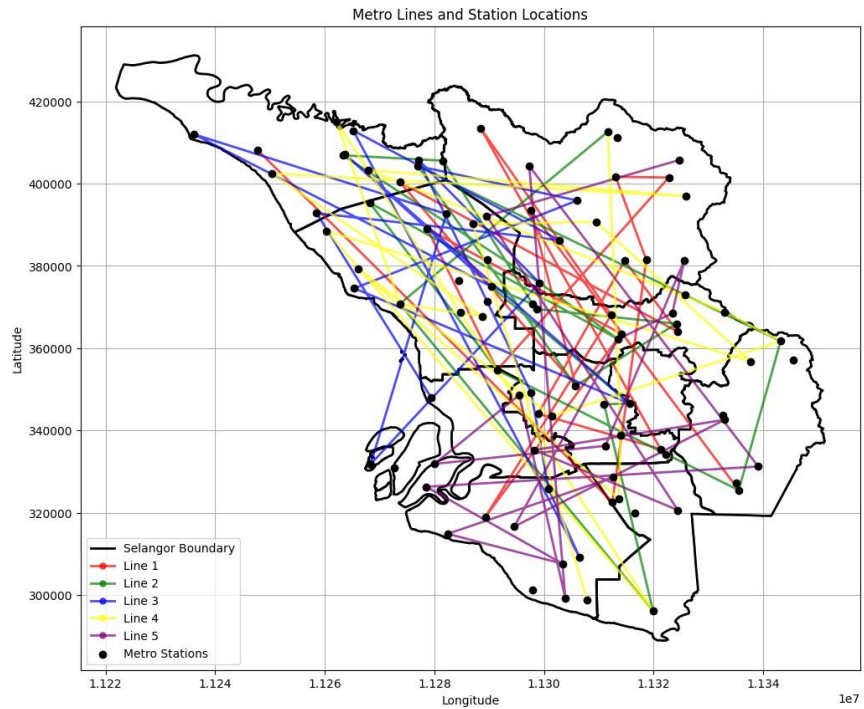

Figure 14: Optimal metro lines configuration.

# 6 Conclusion

In response to the growing demands on Selangor's public transportation system due to rapid urbanization, we aimed to optimize the metro network to better handle increased population density and reduce travel times. Using a genetic algorithm, we developed a framework for determining optimal metro station locations and designing efficient line layouts.

During the course of our study, we encountered several limitations. We were unable to incorporate budget constraints and detailed construction costs into our optimization process. Additionally, due to resource constraints, we had to limit the number of genetic algorithm generations to 10, which may have impacted the optimality of our solutions. Despite these challenges, the framework we created serves as a solid foundation for further development. Future efforts should focus on addressing these limitations and refining the model to incorporate practical considerations for more effective and optimal solutions.

# 7 Acknowledgement

This project was undertaken during a three-month internship at the Faculty of Computer Science and Information Technology, Universiti Malaya, Kuala Lumpur, Malaysia. Special thanks to Universiti Malaya for providing me with this invaluable opportunity to develop and apply skills in the field of urban mobility optimization, also would also like to extend sincere thanks to EMINES – School of Industrial Management, for facilitating this internship and supporting academic and professional growth throughout the project. This internship has not only been an opportunity for academic and professional development but also a rich cultural experience, allowing to immerse in a new environment.

# 8 References

1. Department of Statistics Malaysia. (2024). *Population Table: Administrative Districts*. https://open.dosm.gov.my/data-catalogue/population_district?state=selangor&district=petaling&visual=table

2. Blanco, V., Conde, E., Hinojosa, Y., & Puerto., J. (2020). *An optimization model for line planning and timetabling in automated urban metro subway networks. A case study*. Omega, Volume 92, 102165

3. Król, A., & Król. M., (2019). *The Design of a Metro Network Using a Genetic Algorithm*. Applied Sciences, 9, no. 3: 433

4. Erban, A. (2021). *The use of the genetic algorithms for optimizing public transport schedules in congested urban areas*. IOP Conference Series: Materials Science and Engineering. 1037 (2021) 012062.

5. Biao, D., Rao, Z., Yin, W., Liu, Y., Fang, J., Wang, Y., & Jin, P. (2024). *The Optimization of Urban Traffic Routes Using an Enhanced Genetic Algorithm: A Case Study of Beijing South Railway Station*. Applied Sciences, 14(14): 6130.

6. Kaleybar, H. J., Davoodi, M., Brenna, M., & Zaninelli, D. (2023). *Applications of Genetic Algorithm and Its Variants in Rail Vehicle Systems: A Bibliometric Analysis and Comprehensive Review*. IEEE Access, VOLUME 11, 68972- 68993